\documentclass[groupedaddress,aip,reprint]{revtex4-1}

\usepackage{graphicx}%
\usepackage{dcolumn}%
\usepackage{bm}%
\usepackage{epstopdf}
\usepackage{amssymb}

\usepackage{units}

\begin{document}

\title{Bifurcation, mode coupling and noise in a nonlinear multimode superconducting microwave resonator}

\author{G. Tancredi}
\email{g.tancredi@rhul.ac.uk}
\author{G. Ithier}
\author{P. J. Meeson}    

\affiliation{Department of Physics, Royal Holloway, University of London, Egham, Surrey, TW20 0EX, UK}

\begin{abstract}
The addition of nonlinearity to an harmonic resonator provides a route to complex dynamical behaviour of resonant modes, including coupling between them. We present a superconducting device that makes use of the nonlinearity of Josephson junctions to introduce a controlled, tunable, nonlinear inductance to a thin film coplanar waveguide resonator. Considering the device as a potential quantum optical component in the microwave regime, we create a sensitive bifurcation amplifier and then demonstrate spectroscopy of other resonant modes via the intermode coupling. We find that the sensitivity of the device approaches within a factor two quantitative agreement with a quantum model by Dykman, but is limited by a noise that has its source(s) on-chip.
\end{abstract}

\maketitle
Superconducting circuits that involve the quantisation of charge or flux have been the subject of great attention in recent years as potential building blocks for quantum computers\cite{Wallraff2004,Chiorescu2004,Blais2004}. In this context nonlinear effects have become a strong focus because of unprecedentedly strong coupling parameters, tunability and engineering flexibility. In the weak nonlinear regime the effects are classified into two types: Self-Kerr effects, which give rise to nonlinear phase shifts in a single mode, have been used to generate squeezed states~\cite{Castellano2008}, for parametric amplification~\cite{Castellano2007,Vijay2011} and to perform high-fidelity measurements of quantum bits \cite{Vijay2011,Metcalfe2007,Mallet2009,Ong2011}, while cross-Kerr effects, which dispersively couple two distinct modes, have been discussed in the context of QND measurements and intermode coupling~\cite{Buks2006,Kumar2010,Suchoi2010}. Nonlinearities have also been exploited in other systems~\cite{Ong2013,Kirchmair2013,Kippenberg2004,Westra2010,Schmidt1996,Hau1999, Kang2003,Mucke2010}. 
\begin{figure}
\begin{centering}
\includegraphics[trim=3cm 4.3cm 0cm 3.8cm,clip,scale=0.60]{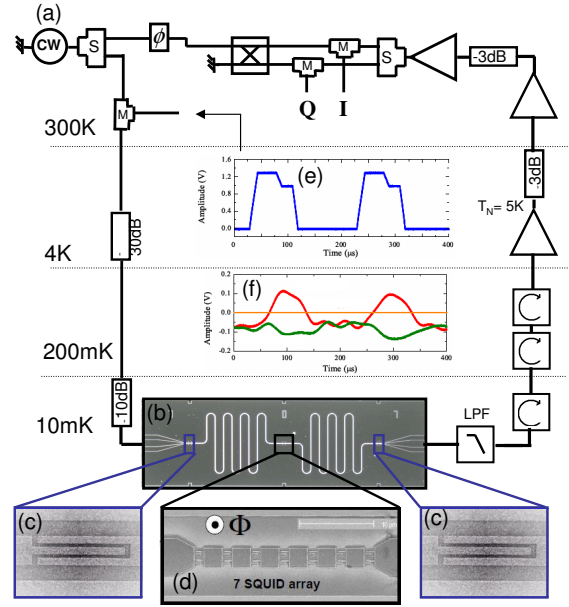}
\caption{\label{fig:sample} \textbf{a}: experimental arrangement for pulsed measurements of a non-linear superconducting resonator. RF pulses with the envelope shown in insert \textbf{e} are fed to the sample (insert \textbf{b}) located in a dilution refrigerator. The power and the frequency of the pulses are tuned so that the bifurcation regime of the nonlinear resonator is probed. The transmitted signal is homodyne detected and recorded, insert \textbf{f} shows the signal obtained when the sample does or does not exhibit switching to the high amplitude metastable state. Inserts \textbf{c} and \textbf{d} show the input and output capacitance and the SQUID array respectively.}
\par\end{centering}
\end{figure}

In this paper, we take a slightly different viewpoint for these systems by considering them as controllable (quantum) optical devices in the microwave range. The development of ``on-chip superconducting microwave optical components'' (SMOCs) such as phase shifters, mixers, splitters and detectors, especially in the single and few photon regime, will be essential for any future superconducting QIP technology. However, despite strong recent advances, non-linear quantum dynamics is relatively poorly understood and exploited and we study here a relatively simple generic SMOC device. Superconducting coplanar resonators, that may be thought of as analogous to optical Fabry-Perot resonators in terms of mode structure, exhibit weak intrinsic Kerr non-linearities. However, the strength of these nonlinearities may be greatly enhanced and controlled by inserting nonlinear, non-dissipative lumped elements based on Josephson junctions\cite{Boaknin2007}. For example, inserting a SQUID in the central conductor couples the SQUID itself to any harmonic mode that has a non-zero current at the SQUID's location\cite{Palacios2008}. It also adds an additional inductance whose value depends on the total current through the junctions which may arise from both externally applied flux and the presence of photons in any coupled harmonic modes. This dependence induces nonlinearities in the coupled modes as well as couplings between these modes. The SQUID also enables flux-tunability of the frequencies of the coupled harmonic modes, together with enhancement and tunability of the Kerr nonlinearities.

We explore Kerr effects in a capacitively coupled, multimode superconducting coplanar resonator which incorporates a seven-element SQUID array located at the mid-point in the central conductor of the resonator\cite{Palacios2008}. The resonator exhibits electromagnetic standing-wave modes with a current node at the location of the capacitors. With the addition of the SQUID array at the cavity mid-point, antinodes of current occur at the location of the SQUIDs for odd-numbered modes. The even-numbered harmonics, in contrast, possess nodes of electrical current at the location of the SQUIDs and do not exhibit enhanced nonlinearities. We first use the self-Kerr effect to operate an odd-numbered mode as a cavity bifurcation amplifier\cite{Boaknin2007}. We then use the sensitivity of this mechanism in conjunction with the cross-Kerr coupling to demonstrate spectroscopy of other cavity modes. We observe that our device is affected by a source of parametrically coupled noise at low frequency; elimination of such effects are a grand challenge in this field~\cite{VanHarlingen2004,Bialczack2007,Anton2012} and so we investigate some of its properties. We then compare our results to the calculation by Dykman~\cite{Dykman1988}, and show that our device operates only a factor two worse than this theory predicts. More details and measurements on this sample can be found in Palacios-Laloy\textit{et al.}\cite{Palacios2008}.

Our device consists of a $200\,$nm sputter-deposited Nb film fabricated through optical lithography into a meandering half-wavelength coplanar waveguide resonator with end capacitors of design value $C=7\,$pF (Figure~\ref{fig:sample}). The fundamental frequency is $\simeq1.77\,$GHz. An array of $N=7$ $Al/AlO_{x}/Al$ SQUIDs was formed  with standard e-beam lithography and double-angle shadow evaporation. The areal dispersion of the SQUID array is $4\%$, according to SEM imaging. The purpose of using a short array is to be able to vary the critical current and hence the magnitude of the nonlinearities at the design stage, while maintaining a fixed overall inductance of the array. The sample is magnetically and electrically shielded and attached to the mixing chamber of a dilution refrigerator. The measurement circuit is shown in Fig.~\ref{fig:sample}. Carefully controlled microwave pulses are transmitted through the sample using lines designed to avoid thermal and amplifier noise from reaching the sample. The pulses are homodyne detected and measured using an oscilloscope or analogue-to-digital converters. We use a pulse repetition rate ($5\,$kHz) that is sufficiently slow that the mode relaxes to its ground state between pulses. For CW measurements an Anritsu Vector Network Analyzer is used. 
The measurement bandwidth is limited by the circulators to approximately $4-8\,$GHz, covering the $2^{nd}$, $3^{rd}$ and $4^{th}$ modes. The third mode is used as our bifurcation amplifier the $2^{nd}$ and $4^{th}$ modes are uncoupled and can be ignored.

We write the Hamiltonian of the system, consisting of the $3^{rd}$ mode and some other odd
 $n^{th}$ mode to which it couples, as $H=H_{n}+H_{3}+H_{3,n}$, with:
\begin{equation}
H_{n}=h\nu_{n}\left(a_{n}^{\dagger}a_{n}+1/2\right),
\end{equation}
\begin{equation}
H_{3}=h\nu_{3}\left(a_{3}^{\dagger}a_{3}+1/2\right)+hK_{3}\left(a_{3}^{\dagger}a_{3}\right)^{2},
\end{equation}
\begin{equation}
H_{3,n}=h\lambda_{3,n}a_{3}^{\dagger}a_{3}a_{n}^{\dagger}a_{n}.
\end{equation}
$a_{n}(a_{n}^{\dagger})$ represents the annihilation (creation)
operator of the $n^{th}$ mode of resonant frequency $\nu_{n}$. $K_{n}$ is the self-Kerr parameter of the $n^{th}$ mode and $\lambda_{n,m}$ is the cross-Kerr coefficient between the $n^{th}$ and $m^{th}$ modes. Here we assume that all self-Kerr terms except for $K_{3}$ may be neglected, as these modes are either operated at low power far from bifurcation or are empty. $K_{3}$, the self-Kerr coefficient of the $3^{rd}$ harmonic, is related to the critical number of photons $N_{3}^{c}$ required
for bifurcation via $N_{3}^{c}=2\,\gamma_{3}/\sqrt{3}K_{3}$,
with $\gamma_{3}$ being the linewidth of the mode. $\lambda_{3,n}$, the
cross-Kerr parameter, defines the frequency shift of the $3^{rd}$
mode per photon present in the $n^{th}$ coupled mode. 

The Kerr coefficients are related to the circuit parameters, and to each other, by
\begin{equation}\label{Kerr-param}
\frac{K_{3}}{\nu_{3}}=\frac{\lambda_{n,3}}{\nu_{n}}=\frac{\beta^{2}}{N}\:\frac{h\nu_{3}}{E_{j}},
\end{equation}
where $E_{j}$ is the Josephson energy of a single SQUID of inductance
$L_{j}$ and $N$ is the number of SQUIDs in the array. $\beta=L_{array}/L_{tot}$ is the ratio between the total
inductance of the SQUID array ($L_{array}=NL_{j}$) and the total
inductance of the resonator ($L_{tot}=L_{wg}+L_{array}$, where $L_{wg}$ is
the inductance of the waveguide). 
Both self- and cross-Kerr coupling parameters in this device are flux-tunable, though not independently. This arises from the fact that $\beta$, $E_{j}$, $\nu_{3}$ and $\nu_{n}$ can
all be tuned by varying the magnetic flux through the SQUIDs. Remarkably, Eq.(\ref{Kerr-param}) has an upper limit given by $ \approx 2 \pi \frac{Z_0}{R_k} \approx 0.01$ where $Z_0$ is the characteristic impedance of the waveguide and $R_k$ is the resistance quantum.

\begin{figure}
\begin{centering}
\includegraphics[trim=1.5cm 0cm 1.5cm 1.5cm,clip,scale=0.30]{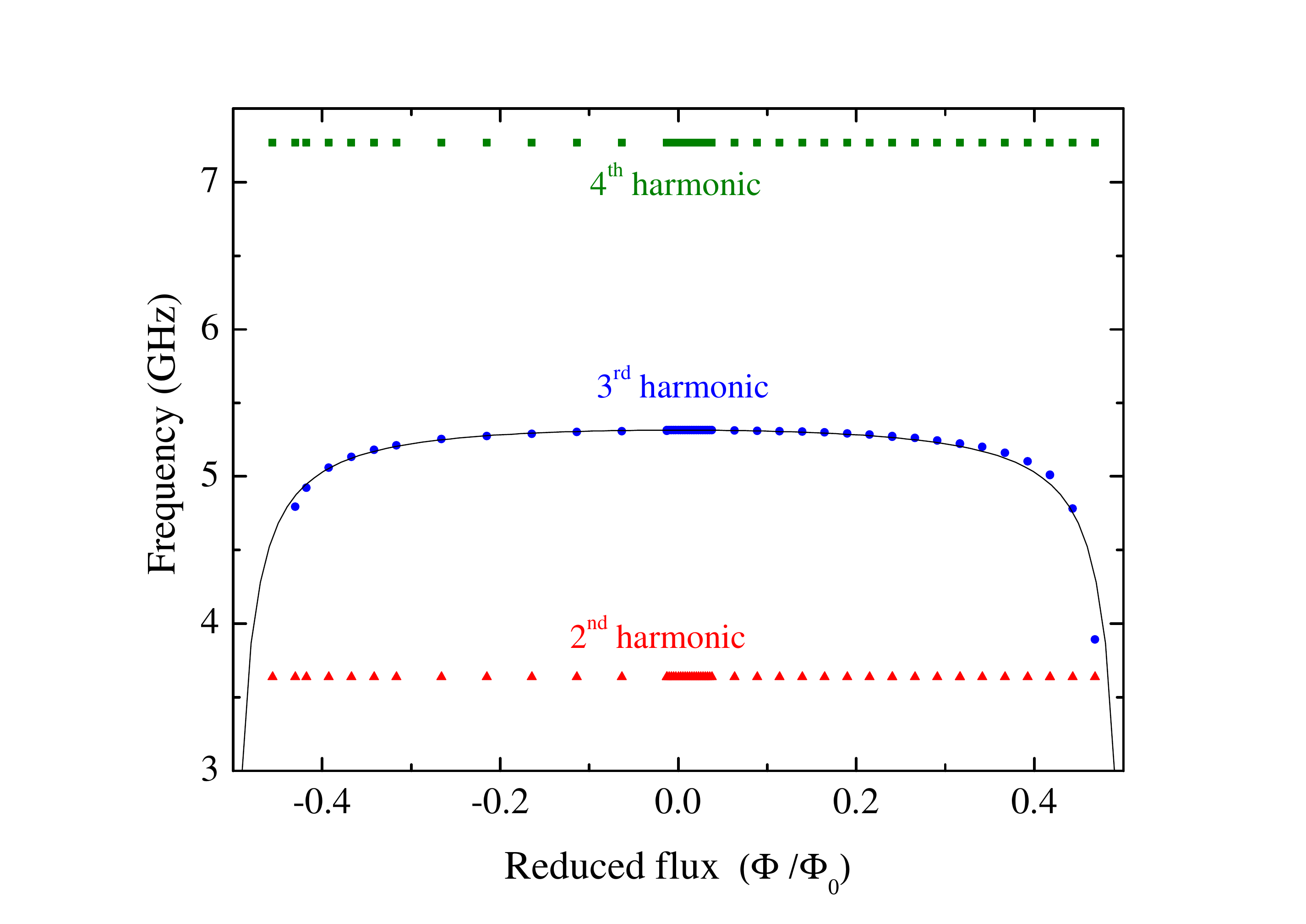}
\caption{\label{fig:tunability}The resonant frequency of the $2^{nd}$, $3^{rd}$ and $4^{th}$ harmonic modes as a function of reduced magnetic flux. Each point is extracted by a Lorentzian fit of the measured  $S_{21}$ forward-scattering parameter. }
\par\end{centering}
\end{figure}

To extract some basic parameters of the system, we used continuous-wave measurements to acquire the forward-scattering parameter, $S_{21}$, for the $2^{nd}$, $3^{rd}$ and $4^{th}$ 
modes at different values of magnetic flux. Figure~\ref{fig:tunability} shows the resonant frequencies as a function of reduced magnetic flux. As expected, the flux modulations of the uncoupled modes $\nu_{2}$ and $\nu_{4}$ are negligible, while $\nu_{3}$ shows a modulation of $\sim30\%$. We obtain the following physical parameters at $\Phi/\Phi_{0}=0$:
$L_{array}=(0.34\pm0.02)\,$nH, the average critical current of a SQUID $I_{c}\simeq6.72\,\mu$A,
$\beta=(2.54\pm0.02)\%$, $\nu_{3}\simeq5.32\,$GHz and $K_{3}\simeq940\,$Hz. Hence $K_{3}/\nu_{3}\simeq2\times 10^{-7}\,$, which, given the linewidth of the mode in the linear regime $\gamma_{3}\simeq212\,$kHz, establishes that  $N_{3}^{c}$, the critical number of photons required for bifurcation, is $\sim260$, which is in reasonable agreement with the estimated microwave power at the sample.

\begin{figure}
\begin{centering}
\includegraphics[trim=1cm 1.0cm 0.5cm 2cm,clip,scale=0.3]{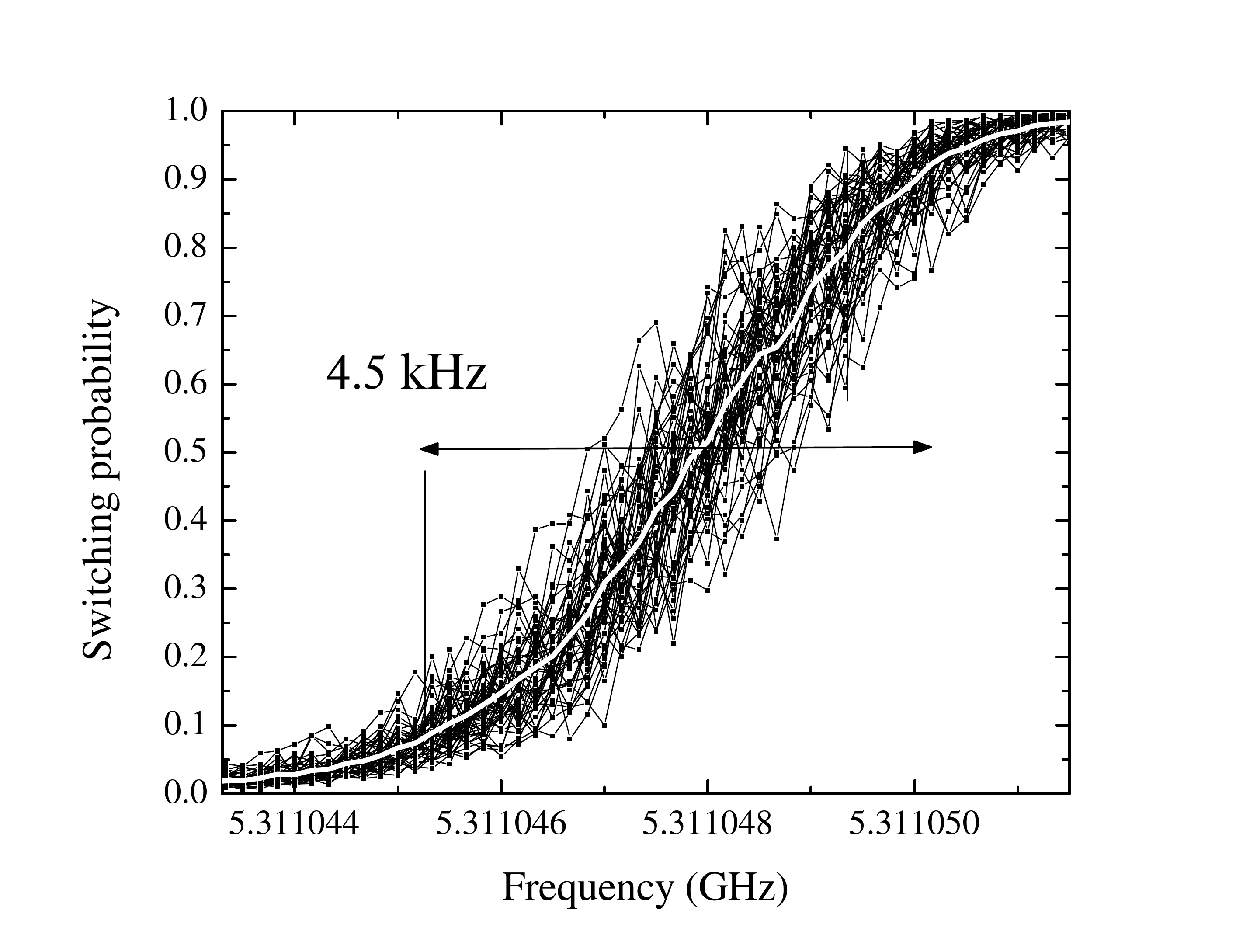}
\par\end{centering}
\caption{\label{fig:S-curves}Measurements of the switching probability of
the $3^{rd}$ mode of a superconducting nonlinear resonator as a function of the driving frequency
at $8$mK and $\Phi/\Phi_{0}=0$. Each point represents the switching probability of the mode as determined from $1000$ pulses. $50$ separate consecutively measured curves are shown. The white line is the average of the $50$ curves.}
\end{figure}

When a cavity mode, in our case the $3^{rd}$, is probed with a number of photons greater than the critical value, the resonator response switches (stochastically and hysteretically) from the low- to the high-amplitude metastable state (see Fig.\ref{fig:sample}\textbf{f}). The probability of the system switching ($P_{S}$) depends weakly on the quality factor of the mode and on the self-Kerr coefficient $K_{3}$ while it strongly depends on the power and frequency of the probing pulses, and on the resonant frequency $\nu_{3}$ of the mode in the linear regime\cite{Dykman1979}. Figure~\ref{fig:S-curves} shows $50$ switching probability curves as a function of the driving frequency $\nu_{d}$. The 10\%--90\% width of the S-shaped averaged probability curve (white line) is $\Delta S=(4.5\,\pm\,0.5)\,$kHz
or $\Delta S/\nu_{3}\simeq0.9\, $ppm. This value is in reasonable agreement
with the theoretical value  calculated from Dykman's model\cite{Dykman1988} ($\Delta S/\nu_{3}\simeq0.5\,$ppm) given by
\begin{equation}
\frac{\Delta S}{\nu_{3}}=\frac{3^{\nicefrac{2}{3}}}{4}\,\left(\frac{\beta^{2}}{N}\right)^{\nicefrac{2}{3}}\left(\frac{k_{B}T_{eff}}{E_{j}}\right)^{\nicefrac{2}{3}}\left(\frac{\delta\nu}{\nu_{3}}\right)^{\nicefrac{1}{3}},
\label{eq:DeltaS}
\end{equation}
where $\delta\nu=\nu_{d}-\nu_{3}$ is the frequency detuning of the microwave
driving and $T_{eff}=\frac{h\nu_3}{2k_b} \coth{\frac{h\nu_3}{2k_b T}}$  is the effective temperature. The effective temperature $T_{eff}$ is the physical temperature if $T>T_{co}$ or $h\nu_3/2k_b$ for $T \ll T_{co}$, where $T_{co}=h\nu_3/4k_b\simeq70\,$mK is the crossover temperature from the classical regime to the quantum regime~\cite{Dykman1988}. For comparison, the linewidth of the same mode in the linear regime is $\gamma_{3}=212\,$kHz or $\gamma_{3}/\nu_{3}=40\,$ppm. $\Delta S$ represents the sensitivity of the device - i.e. it expresses the ability to resolve a frequency shift of $\Delta S$ with a high level of confidence within a single shot measurement.
This bifurcation technique, known as cavity bifurcation amplifier\cite{Boaknin2007}, is a sensitive threshold measurement which may now be used to investigate other aspects of the resonator.

\begin{figure}
\begin{centering}
\includegraphics[trim=0.4cm 0.6cm 0cm 0.5cm,clip,scale=0.33]{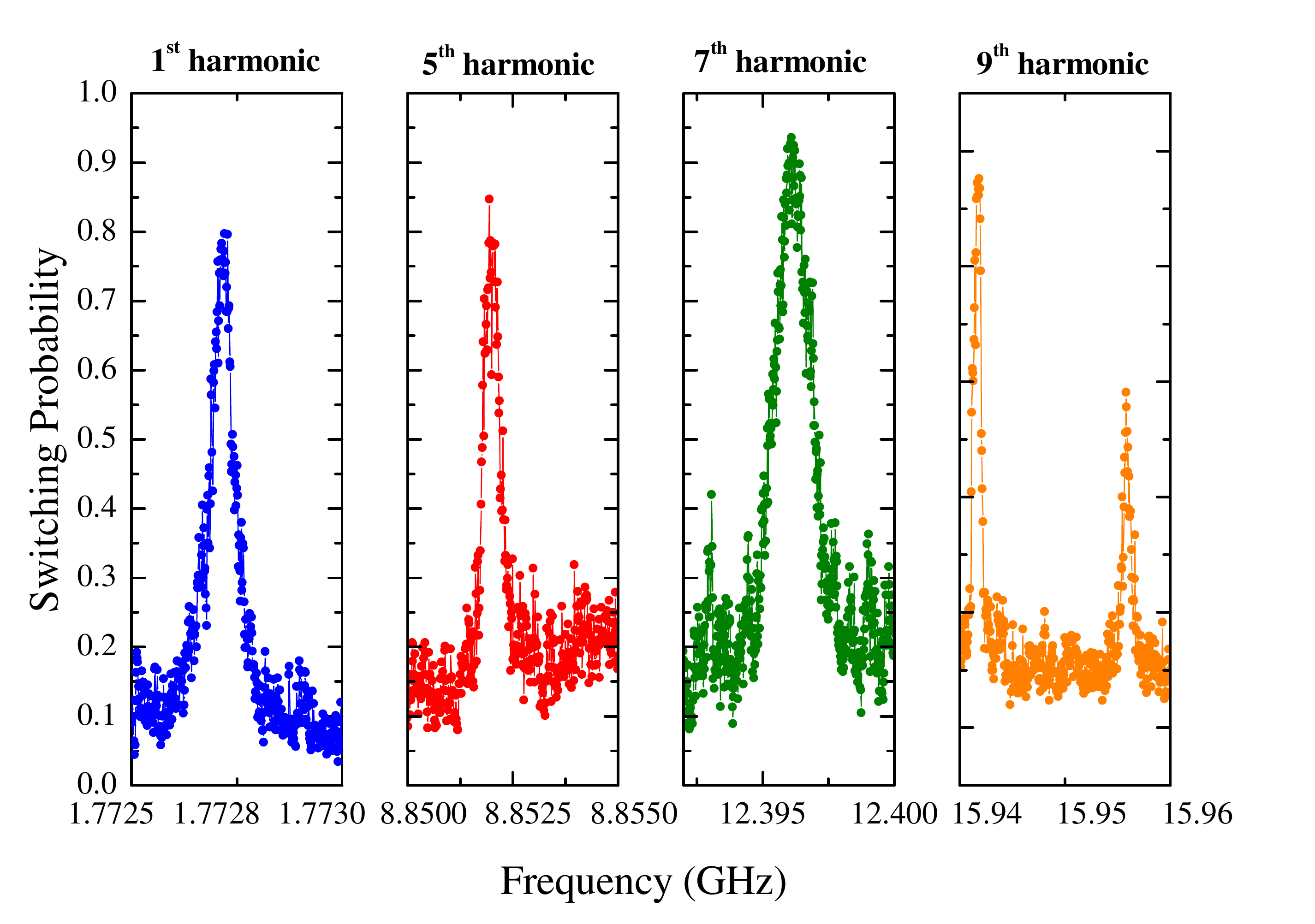}
\par\end{centering}
\caption{\label{fig:spectroscopy}  Spectroscopy of Kerr-coupled microwave modes. The
switching probability of the $3^{rd}$ mode is measured while
simultaneously applying a continuous swept microwave excitation to the $1^{st}$,
$5^{th}$, $7^{th}$, and $9^{th}$ modes. }
\end{figure}

To demonstrate the potential utility of this SMOC, we performed spectroscopy of nearby coupled modes. We first tuned the bifurcation amplifier to a switching probability of  $10\%$ and then injected a low-power continuous-wave signal and swept its frequency in the vicinity of the coupled mode to be detected. As shown in Fig.~\ref{fig:spectroscopy}, we were able to detect the change in photon occupation of the $1^{st}$, $5^{th}$, $7^{th}$ and $9^{th}$ modes via a variation in the switching probability of the $3^{rd}$ mode, thus demonstrating the cross-Kerr coupling between the modes. The resonant frequencies
of the coupled modes are obtained via a Lorentzian fit providing a value for $\beta$ at zero magnetic flux of $\beta=(2.55 \pm 0.1)\%$ that is compatible with 
the previously extracted value. In this way, the cross-Kerr effect enables spectroscopy of modes that lie outside our measurement bandwidth and thus are undetectable via the usual transmission detection. Based on the estimate $\frac{\Delta S}{\lambda_{1,3}}\simeq 10$, our device is sensitive to $\simeq10$ photons in the $1^{st}$ mode. The $9^{th}$ mode is split for reasons that we do not understand; possibly the mode is bifurcating.

We now consider in more details the sensitivity of our device. We have already noted that Dykman's calculation\cite{Dykman1988} predicts a width $\Delta S$ due to quantum fluctuations that is below our experimental value. The probability curves repetitively acquired under nominally identical conditions (Fig.~\ref{fig:S-curves}) show that the switching probability $P_{S}$ is affected by fluctuations that significantly exceed the expected statistical fluctuations. A noise source, whose low frequency components appear in these measurements, may also be present at higher frequencies where it may increase  $\Delta S$ and be the source of the discrepancy between our experimental values and the theory of Dykman. Such a noise source would be parametrically coupled to our bifurcation amplifier, either through the frequency of the cavity or through the amplitude and frequency of the biasing pulses. Estimates of the extrinsic sources of noises, such as temperature fluctuations and power (frequency) fluctuations of the biasing pulses, suggest that they cannot account for the observed fluctuations in $P_{S}$. Fluctuations in the microwave pulse power at the sample holder, measured with the cryostat warm, were found to contribute  $\lesssim 10\%$ to the observed $\Delta S$. Hence, in the following, we focus our attention on noise sources intrinsic to the device. 

\begin{figure}
\begin{centering}
\includegraphics[trim=1.5cm 0cm 0cm 0cm,clip,scale=0.35]{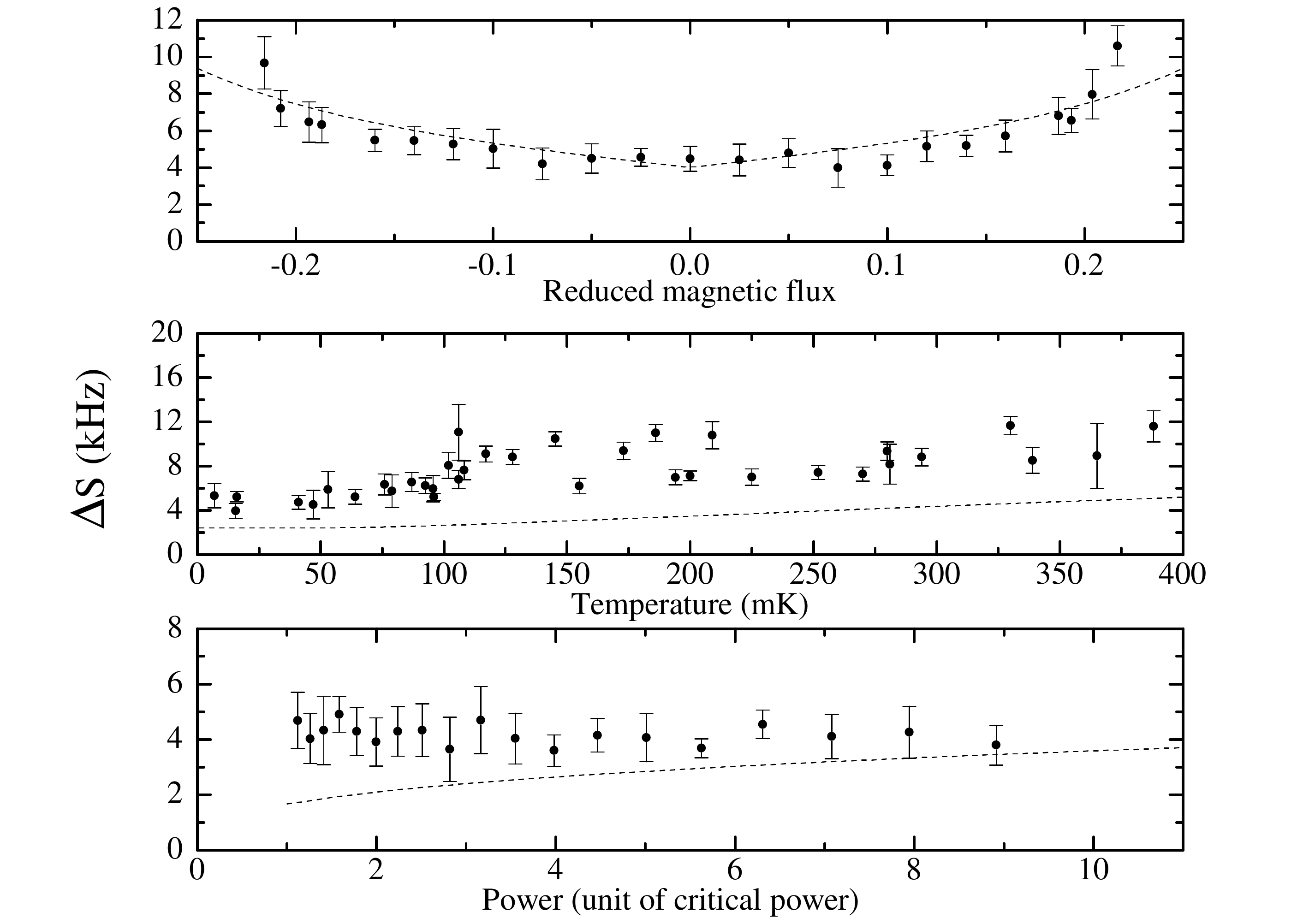}
\par\end{centering}
\caption{\label{fig:width} The width $\Delta S$ of the switching probability curve as a function of reduced magnetic flux (top panel), temperature (central panel), and driving microwave power (lower panel). In the upper and lower panel the data was taken at refrigerator base temperature $\approx 8$mK. For the center panel and lower panels the data was taken at zero magnetic flux and the dashed lines represent the prediction of Dykman's theory. }
\end{figure}

The upper panel in Fig.~\ref{fig:width} shows the dependence of the width $\Delta S$
on applied magnetic flux. We first note that $\Delta S$ displays a minimum of $\Delta S_{0}\simeq4.5\,$kHz. We achieve a reasonable fit by assuming that $\Delta S_{0}$ is an additive constant at all flux values and that the additional increase of $\Delta S$ away from zero flux is caused by a quasi-static flux noise. The RMS amplitude of the flux noise, integrated over the bandwidth of the experiment, can be extracted using the measured dependence of $\nu_{3}$ on magnetic flux. This leads to an RMS noise amplitude of $\simeq5\,\mu\Phi_{0}$, which is comparable to commercial SQUID amplifiers and probably indicates the quality of our magnetic shielding. The second-order contribution to $\Delta S$ from this flux noise at zero flux is  $\simeq1$Hz, which is an insignificant contribution to the observed $\Delta S_{0}=4.5\,$kHz. Hence, $\Delta S$ at zero flux is not due to flux noise. 
The center panel in Fig.~\ref{fig:width} shows measurements of $\Delta S$ as a function of the refrigerator temperature over the range $8-400\,$mK at zero magnetic flux. The expected dependence of $\Delta S$ with temperature is given by  Eq(\ref{eq:DeltaS}), with  $T_{eff}=\frac{h\nu_3}{2k_b} \coth{\frac{h\nu_3}{2k_b T}}$ (shown as the dotted line in the central panel of Fig.~\ref{fig:width}). We observe that the values for $\Delta S$ are generally within a factor two of Dykman's predictions and follow the weak temperature dependence of the theory down to the lowest temperature.
The lower panel in Fig.~\ref{fig:width} shows $\Delta S$ as a function of the microwave power in units of the critical power for bifurcation. $\Delta S$ is found to be almost constant over one order of magnitude above the critical power and is again within a factor two of the predictions of Dykman's model.

In summary, we have created a device to assess the usefulness of on-chip superconducting microwave optical components, or ``SMOCs''. By inserting a controllable, nonlinear element into a superconducting microwave resonator, we are able to create a SMOC that simultaneously exploits the self- and cross-Kerr effects. We have shown that high-sensitivity spectroscopy of nearby modes may be performed through measurements of the probability of switching of another mode via non-linear coupling. Analysis of the device sensitivity suggests that there is a source of noise that manifests as fluctuations in the cavity resonant frequency, it has a low frequency component and it arises on-chip. The noise may be related to the ubiquitous noise seen in similar SQUID and qubit systems; we have excluded microscopic flux noise and the remaining sources include two-level fluctuators in the environment and critical current noise in the tunnel barriers. However, we note that the intrinsic sensitivity of the device is close to the limits given by the theory of Dykman.

\begin{acknowledgments}
We wish to thank EPSRC and the Leverhulme Trust for their financial support; D. Esteve and all at the Quantronics Group at CEA Saclay for their support, especially P. Bertet who fabricated the sample, A. Tzalenchuk, T. Lindstrom (NPL) for helpful discussions and the loan of equipment, and N. K. Langford for helpuful discussions and a critical reading of the manuscript.
\end{acknowledgments}

\end{document}